\documentclass{cimento}

\newcommand{\be}{\begin{equation}}
\newcommand{\ee}{\end{equation}}
\newcommand{\ba}{\begin{eqnarray}}
\newcommand{\ea}{\end{eqnarray}}
\newcommand{\beg}{\begin{gather*}}
\newcommand{\eng}{\end{gather*}}
\newcommand{\hh}{,\hspace{0.5cm}}
\newcommand{\hhh}{,\hspace{0.2cm}}
\newcommand{\BE}{\begin{eqnletter}}
\newcommand{\EE}{\end{eqnletter}}

\newcommand{\eq}[1]{(\ref{#1})}

\newcommand{\n}[1]{\label{#1}}
\newcommand{\ts}[1]{{\boldsymbol{#1}}}

\newcommand{\CAL}{\mathcal}

\newcommand{\inds}[1]{{\scriptscriptstyle #1}}
\newcommand{\ins}[1]{{\mbox{\tiny #1}}}

\usepackage{graphicx}  
\usepackage{subcaption}
\usepackage{amsmath,latexsym}

\title{Limiting curvature models of gravity}
\author{Valeri P.~Frolov\from{ins:x}}
\instlist{\inst{ins:x} Theoretical Physics Institute, University of Alberta, Edmonton, Alberta, Canada T6G 2E1}

\begin{document}

\maketitle

\begin{abstract}
In this paper we discuss a limiting curvature condition and its application to cosmological and black hole solutions.
\end{abstract}

\section{Introduction}

Existence of singularities is a generic property of general relativity. For well known cosmological and black-hole  solutions of Einstein equations these singularities are related to infinite growth of curvature. One can either ``live" with  a situation when the spacetime terminates its existence at the singularity, or believe that this ``decease" of general relativity can be cured by proper modification of Einstein equations in the ultraviolet
domain. There exists a lot of publications where different modifications of general relativity were proposed. At the same time one may try to answer a ``simpler question": What might be properties of cosmological and black-hole models in a theory which forbids the curvature singularity formation.

In 1982 Markov proposed a so called {\it limiting curvature condition} \cite{MA1,MA2}. According to this conjecture the spacetime curvature should be always restricted by some universal value
\be \n{RRR}
|{\cal R}|<\Lambda={B_{\cal R}\over \ell^2}\, .
\ee
Here ${\cal R}$ is a scalar curvature invariant which has dimension of $[length]^{-2}$ and $\Lambda$ is its limiting value. The parameter $\ell$ connected with the radius of the curvature plays a role of a fundamental length of the theory. The dimensionless parameter $B_{\cal R}$ depends on a choice of the curvature invariant ${\cal R}$ but it is universal in the following sense. It does not depend on the parameters which enter solutions, such as black hole mass or the total entropy of the universe.

There are a lot of publications where nonsingular cosmological and  black hole models are discussed. In principle, there exist two kind of approaches to this problem
\begin{itemize}
\item One can try to guess the form of the metric which satisfies the limiting curvature condition and which in the spacetime domain where the curvature is much smaller than the critical one properly reproduces corresponding solutions of Einstein equations;
\item One can try to find such a modification of the Einstein equations which automatically prevents infinite growth of the curvature.
\end{itemize}
In the present paper we discuss some aspects of these two approaches.

\section{Metrics satisfying the limiting curvature condition}

Let us first describe an interesting class of metrics which satisfy the limiting curvature condition. Let us consider the following metric
\BE\n{metric}
&&ds^2=-f(r,v) dv^2+2 dv dr +r^2 d\omega^2\, ,\\
&& f(r,v)=1-{m^2(v)\over \ell^2}F(x)\hh x= {r\over m(v)} . \n{ff}
\EE
Here $m(v)$ and $F(x)$ are two arbitrary functions and $d\omega^2$ is a metric on a unit round $2D$ sphere.
The parameter $\ell$ which enters this metric has the dimension of length. At the moment it is arbitrary. Later $\ell$  will be identified with the radius corresponding to the limiting curvature.
The mass function $m$ controls the mass of the black hole and when $dm/dv\ne 0$ it is related to the energy density of the spherically symmetric null fluid flux, which changes the black hole parameters.
We shall demonstrate that for a proper choice of structure function $F(x)$ metric (\ref{metric}) describes a nonsingular spherically symmetric black hole and its spacetime obeys the limiting curvature condition.

A well known special case of (\ref{metric}) is Vaidya metric
\be  \n{VAD}
ds_V^2=-\left(1-{2M(v)\over r}\right)dv^2+2 dv dr +r^2 d\omega^2\, .
\ee
It describes a black hole. Here $M(v)$ is the mass of the black hole as a function of the advanced time $v$. For constant $M$ metric (\ref{VAD}) coincides with Schwarzschild metric. If $dM/dv\ne 0$ there exists a null fluid flux which changes the black-hole mass. Vaidya metric (\ref{VAD}) is often used to describe a formation and consequent evaporation of a black hole. At the phase of formation $dM/dv> 0$, while during the evaporation phase $dM/dv<0$.
It is easy to check that metric  (\ref{VAD}) can  be presented in the form (\ref{metric}) with the following choice of the mass function
\be
m(v)=[2M(v)\ell^2]^{1/3}\, .
\ee

Our next step is calculation of the curvature invariants for metric (\ref{metric}).
We use the following notations
\BE
&&C^2=C_{\alpha\beta\gamma\delta}C^{\alpha\beta\gamma\delta}\hhh
S^2=S_{\alpha\beta}S^{\alpha\beta}=R_{\alpha\beta}R^{\alpha\beta}-{1\over 4}R^2\,  ,\\
&&K=R_{\alpha\beta\gamma\delta}R^{\alpha\beta\gamma\delta}=C^2+2S^2+{1\over 6}R^2 \, .
\EE
Here $S_{\alpha\beta}=R_{\alpha\beta}-{1\over 4}R g_{\alpha\beta}$,
and $R_{\alpha\beta}$ and $R$ are Ricci tensor and Ricci scalar, respectively.

Calculation of the scalar curvature invariants gives
\begin{eqnletter}
C&=&{1\over \sqrt{3}\ell^2 x^2} (x^2 F''-2xF'+2F)\, ,\n{eqC}\\
S&=&{1\over 2\ell^2 x^2} (x^2 F''-2F)\, ,\n{eqS}\\
R&=&{1\over \ell^2 x^2} (x^2 F''+4xF'+2F)\,   ,\n{eqR}\\
K&=&{1\over \ell^4}\left[  (F'')^2+{4(F')^2\over x^2}+{4F^2\over x^4}
\right] \, .
\end{eqnletter}
These relations demonstrate a remarkable property of the metric (\ref{metric}): The curvature invariants written as functions of $x$ do not depend on the mass function $m(v)$. This means that in order to keep these invariants uniformly and universally bounded it is sufficient to chose a properly bounded function $F(x)$.

Using expressions (\ref{eqC})--(\ref{eqR}) one gets
\be
\ell^2\left( 2\sqrt{3}C-6S+R\right)=12{F(x)\over x^2}\, .
\ee
If the curvature invariants are finite at $x=0$ then the function $F(x)$ at this point should have the following asymptotic form
\be \n{FU}
F(x)=F_0 x^2+\ldots\hh F_0=\mbox{const}\, ,
\ee
where dots denote terms of the higher order in $x$.
Let us write $F(x)$ in the form $F(x)=x^2 U(x)$,  then
\BE
C&=&{1\over \sqrt{3}\ell^2}(x^2 U''+2xU')\, ,\\
S&=&{1\over 2\ell^2}(x^2U''+2xU')\, ,\\
R&=&{1\over \ell^2}(x^2U''+8xU'+12U)\, .
\EE
If the function $U(x)$ and its derivatives at $x=0$ are finite then the scalar curvature invariants are finite at this point. The metric function $f$ at small $r$ has the following form
\be
f=1-U(0){r^2\over \ell^2}+\ldots \, .
\ee
We use an ambiguity in the definition of the parameter $\ell$ to absorb the factor $U(0)$ into it.

In order to have the asymptotically flat metric the function $U(x)$ at large $x$ should have the following form
$U(x)\sim c/x^3 +\ldots$ . Then one has
\be
f\sim 1-{c\ m^3\over \ell^2 r} +\ldots\, .
\ee
The constant $c$ can be absorbed into a redefinition of the mass function $m$. Using this ambiguity we choose $c=1$. For this choice the function $f$ at large distance asymptotically takes the form (\ref{VAD}) if
\be
m(v)=[2M(v)\ell^2]^{1/3}\, .
\ee
Let us summarize. Using the above discussed ambiguities one can impose the following conditions on the function $U(x)$
\be\n{ASU}
U(0)=1\hh U(x\to\infty)\sim 1/x^3\, .
\ee
Using this asymptotic form of $U(x)$ at large $x$ it is easy to show that the leading term of the Einstein tensor $G_{\mu\nu}$ calculated for the metric (\ref{metric}) at large $r$ is
\be
G_{\mu\nu}\approx {2\dot{M}\over r^2}v_{,\mu}v_{,\nu}\, .
\ee

If the following equation
\be\n{app}
x^2 U(x)={\ell^2\over m^2(v)}\, .
\ee
has a solution $x=x(v)$ then the metric (\ref{metric}) has a trapped surface at $x=x(v)$.
Since the function $x^2 U(x)$ grows from $0$ near $x=0$ and decrease as $1/x$ at the infinity it has one or more local maxima. We assume that there exists only one  maximum at some point $x=x_*$. If at this point
\be
x_*^2 U(x_*)>{\ell^2\over m(v)}\, ,
\ee
then equation (\ref{app}) has two solutions, $x_0<x_*(v)$ and $x_1>x_*(v)$. If the mass $M$ is constant and does not depend on advanced time $v$, the outer surface $x=x_1$ is an event horizon and $x=x_0$ is the inner horizon. In a time-dependent case solutions of the equation (\ref{app}) describe positions of the inner and outer branches of the apparent horizon. When the mass $M$ decreases with time then at some time these branches can meet one the other. This happens when $x_0=x_1=x_*(v)$. At this point the apparent horizon terminates. In such a situation the information collected inside the apparent horizon can become visible to an external observer. Technically, this means that the event horizon does not exist.

\begin{figure}[!hbt]
  \centering
  \includegraphics[width=.4\linewidth]{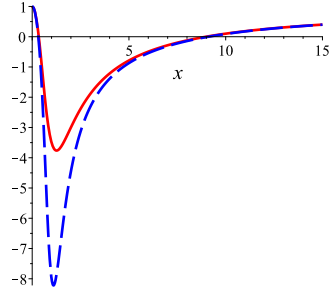}
  \vspace{-2.2cm}
  \caption{Function $f$ for Hayward metric (\ref{UH}) (solid line) and its generalization (\ref{Un}) with $n=2$ (dash line). In both cases the mass $M$ is constant and its value is chosen to be equal $27/2$ in the units where $\ell=1$. A coordinate $x$ along the horizontal axis is $x=r/m=r/3$.
  }
  \label{ff}
\end{figure}

\begin{figure}[!hbt]
\begin{subfigure}{.5\textwidth}
  \centering
  \includegraphics[width=.9\linewidth]{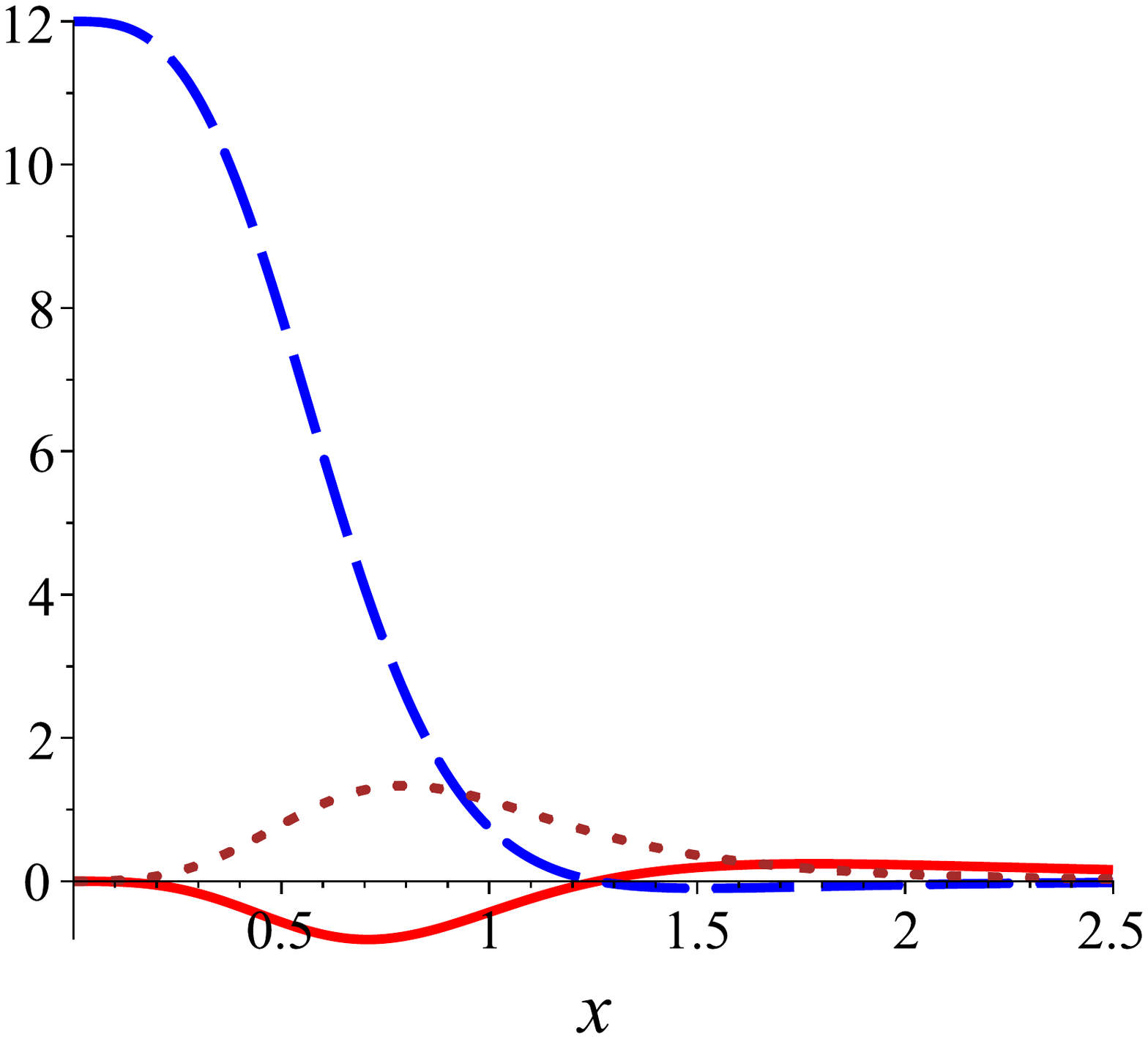}
\end{subfigure}%
\begin{subfigure}{.5\textwidth}
  \centering
  \includegraphics[width=.9\linewidth]{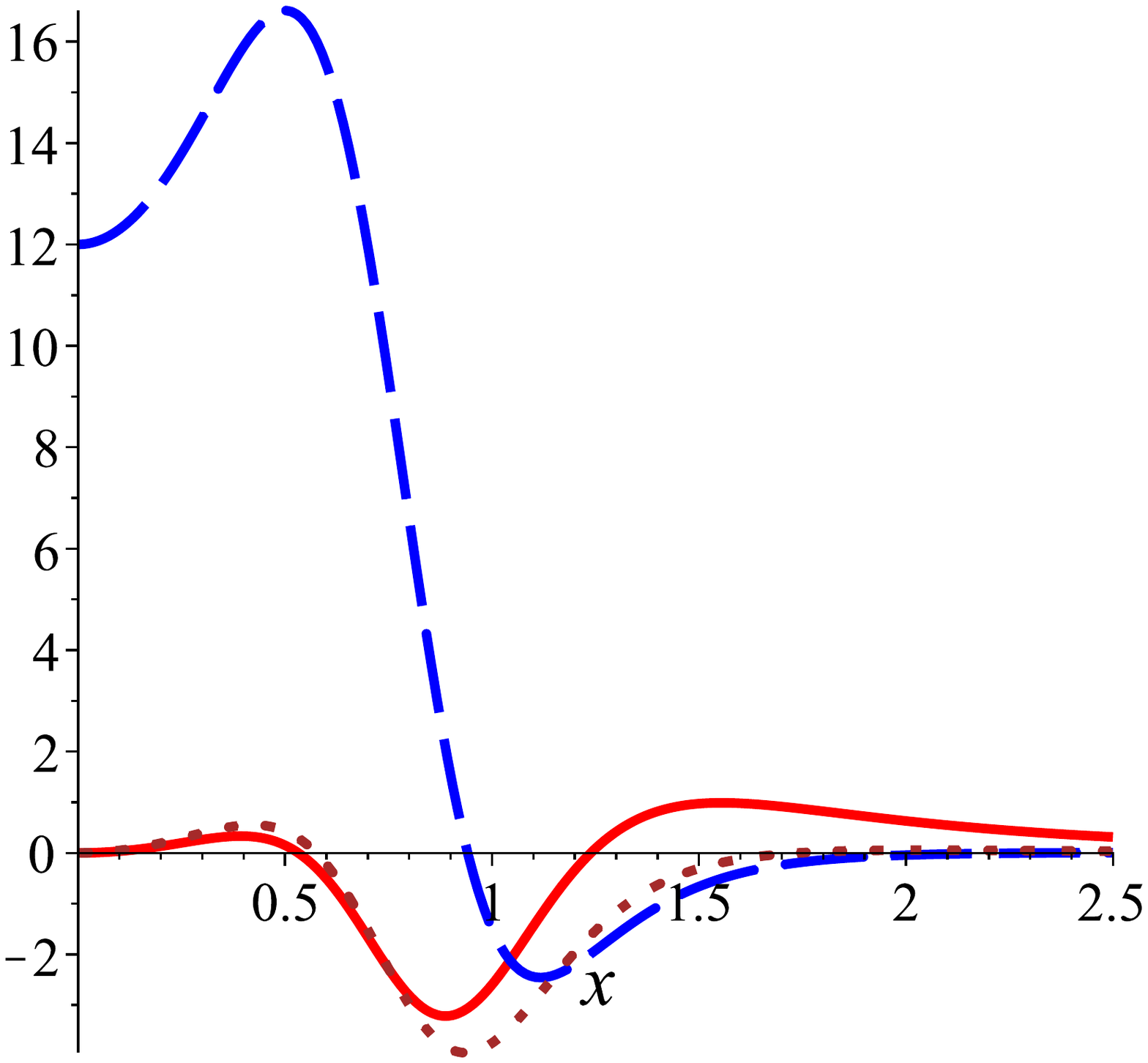}
\end{subfigure}
\vspace{-2.5cm}
\caption{Plots for the curvature invariants for Hayward metric (\ref{UH}) (left) and its generalization (\ref{Un}) (right). The parameters of these metrics are chosen the same as in Fig.~\ref{ff}. Solid, dash and point lines are used for $C$, $S$ and $R$ invariants, respectively.}
\label{CURV}
\end{figure}

The simplest choice of function $U(x)$ satisfying  conditions (\ref{ASU}) is
\be \n{UH}
U={1\over 1+x^3}\, .
\ee
It is easy to check that the function $f(r,v)$ is this case is
\be \n{Hf}
f(r,v)=1-{2M(v)r^2\over r^3+2M(v)\ell^2}\, ,
\ee
and the corresponding metric (\ref{metric}) is nothing but a Vaidya-type generalization of the Hayward metric \cite{Hayward}. Discussion of properties of this metric and its generalizations as well as further references can be found in \cite{F1,B1,F2,B2}.

A solid line in figure~\ref{ff} shows a plot of function $f$ for metric  (\ref{metric}) with $U$ given by (\ref{UH}). For illustration we choose the mass $M$ to be constant and put $2M=27$ and $\ell=1$, so that $m=3$. This plot shows $f$ as a function of $x=r/m=r/3$. The left plot in figure~\ref{CURV} shows invariants $C$, $S$ and $R$ calculated for the same parameters $M$ and $\ell$.

Metrics (\ref{metric}) with function $U(x)$ satisfying conditions (\ref{ASU}) is a far-going generalization of the Hayward-Vaidya metric  (\ref{Hf}). Their common property is that their curvature invariants do not depend on mass function. If $U(x)$ and its derivatives are bounded on the interval $x\in [0,\infty)$, and the following relations are valid
\be\n{MAX}
\max_{x\in[0,\infty)} |U(x)|=U_0\hhh
\max_{x\in[0,\infty)} |xU'(x)|=U_1\hhh
\max_{x\in[0,\infty)} |x^2U''(x)|=U_2\, ,
\ee
where $U_0$, $U_1$ and $U_2$ are finite constants,
then the curvature invariants obey the limiting curvature conditions
\be
C^2\le {B_C\over \ell^4}\hhh S^2\le {B_S\over \ell^4}\hhh |R|\le {B_R\over \ell^2}\, .
\ee
Dimensionless constants $B_C$, $B_S$ and $B_R$ depend only on the choice of the function $U(x)$.

There exists infinite number of functions $U(x)$ satisfying both conditions (\ref{ASU})  and (\ref{MAX}). For example one can take
\be\n{Un}
U(x)={1+x^n\over 1+x^{n+3}}\, ,
\ee
where $n$ is a positive integer number. Figure~\ref{ff} shows by a dash line the metric function $f$ for $n=2$.
The curvature invariants for this metric with $n=2$ are shown in the right plots in figure~\ref{CURV}.

\section{Limiting curvature theory of gravity}

Infinitely growing curvature in the black hole interior signals the formation of singularity. Metrics presented in the previous section are nonsingular and they obey the limiting curvature condition. However, they are not solutions of some explicitly known modified gravity equations.
How to modify gravity equations in order to escape singularities? Recently a new approach has been proposed which allows one to prevent infinite growth of the spacetime curvature \cite{FZ1,FZ2,FZ3}. It was called a {\it limiting curvature gravity} (LCG) theory. Its general idea is to incorporate inequality constraints restricting the growth of the curvature into the gravity action.

Suppose we want not to allow some curvature invariant to be arbitrary large. Let us denote this invariant by ${\CAL R}$ and its limiting value by $\Lambda$. We introduce a constraint function
\be
\Phi={\CAL R}-\Lambda\, .
\ee
Thus our goal is to include the following inequality constraint $\Phi\le 0$ into our theory of gravity. This can be done as follows. Let the original (unconstrained) action for the gravitational field be
\be
S_g=\int dx \sqrt{-g} L_g\, .
\ee
$L_g$ is the function of the metric $\bf{g}$ and its derivatives. In the standard general relativity $L_g={1\over 2\kappa}R$, where $R$ is a scalar curvature, and $\kappa=8\pi G$. If there exist other fields beside the gravitational one, then $S_g$  depends on them as well.

We modify this action by adding a term which is responsible for imposed inequality constraint
\BE
&&S=S_g+S_c\, ,\\
&& S_c=\int dx \sqrt{-g} \ \chi\  (\Phi+\zeta^2)\, .
\EE
The extended action $S$ besides the metric $\ts{g}$ contains two Lagrange multipliers, $\chi(x)$ and $\zeta(x)$.
Variation of $S$ with respect to these functions gives
\be\n{CONS}
\Phi+\zeta^2=0\hh \chi\zeta=0\, .
\ee
If $\Phi<0$, then the first equation gives $\zeta^2=-\Phi$, while the second equation implies that $\chi=0$. In this regime, which is called subcritical, the gravitational equations are not modified
\be
{\delta {S}\over \delta g_{\mu\nu}}\bigg|_{\chi=0}={\delta {S}_g\over \delta g_{\mu\nu}}=0\, .
\ee

When the curvature ${\CAL R}$ reaches its critical value $\Lambda$, the function $\zeta$ vanishes. In this so called supercritical regime the second Lagrange multiplier, $\chi$ becomes non-zero. As a result this function enters into the variation of the action $ S$ and modifies the gravitational equations. They contain now $\chi$ and its derivatives. The number of the gravitational equations remain the same, but  they contain one more variable. However, the complete set of equations contains now an additional constraint equation
\be
\Phi=0\, .
\ee
This extended system of equations is sufficient to determine both the metric $\ts{g}$ and the function $\chi$. Since a non-vanishing value of $\chi$ indicates that a system is in its supercritical regime it is called a control function.

Suppose a transition between initial subcritical solution to the supercritical one happens at some junction surface $\Sigma$. One can use obtained system of equations and known subcritical metric in order to find the initial conditions on $\Sigma$ for the corresponding supercritical solution. It is sufficient to use the continuity conditions which follow from the field equations.

The extended action $S$ can be easily generalize to the case of several inequality constraints. For this purpose it is sufficient to substitute instead of $\chi\  (\Phi+\zeta^2)$ in the integrand in $S_c$ a sum
\be
\sum_i \chi_i\  (\Phi_i+\zeta_i^2)\, .
\ee
Here $\Phi_i$ is a constraint function for the constraint number $i$, and $\chi_i$ and $\zeta_i$ are two Lagrange multipliers accompanying this constraint.

\section{Two-dimensional black holes in LCG theory}

To illustrate how a described in the previous section approach works let us consider a  $2D$ black hole LCG model \cite{FZ1}. We start with the action of a 2D dilaton gravity model which is known to have black-hole solutions and is exactly solvable at the classical level
\ba\label{Witten1}
S_g=\frac{1}{2}s
    \int d^2x~ |g|^{1/2}~\Big(\psi^2 R+4(\nabla\phi)^2+4\lambda^2\Big).
\ea
Here $R$ is the curvature of the two-dimensional spacetime with metric $g_{\alpha\beta}$ and $\psi$ is a dilaton field. Parameter $\lambda$ has the dimension $[length]^{-2}$.
This model naturally appears in the framework of string theory and its properties have been extensively studied.

A limiting curvature model is obtained  by adding to $ S_g$ an action
\be \label{I1}
S_c=\frac{1}{2}\int d^2x~ |g|^{1/2}~\chi~\big( \Phi+\zeta^2\big)\hh
\Phi=R-\Lambda\, .
\ee
Here $\Lambda>0$ is a positive constant constraining the  $2D$ curvature $R$.
Without the Lagrange multiplier $\zeta$ the action \eq{I1} looks exactly as that of  Jackiw-Teitelboim gravity model.

Variation of the action $S=S_g+S_c$ over $\zeta$ and $\chi$ gives the constraint equations
(\ref{CONS}), while its
variation   over $\psi$ and $g_{\alpha\beta}$ leads to the dilaton and gravitational field equations
\BE\label{eqpsi1}
&&\Box\psi-\Big(\lambda^2+\frac{1}{4}(\Lambda-\zeta^2)\Big)\psi=0, ,\\
&&\chi_{;\alpha\beta}+\frac{1}{2}g_{\alpha\beta}R\chi=Q_{\alpha\beta}\, ,\\
&&Q_{\alpha\beta}=4\psi_{;\alpha}\psi_{;\beta}
\!+\!\frac{1}{2}g_{\alpha\beta}\big[\!-\! 4\psi_{;\epsilon}\psi^{;\epsilon}\!+\!(4\lambda^2+R)\psi^2\big].\label{QQ}
\EE

In the subcritical domain these equations coincide with the standard equations of $2D$ dilaton gravity and their solutions are well known.
The solution of the field equations  (\ref{eqpsi1})   is
\BE \label{Schw}
&&ds^2=-f dt^2+f^{-1}dr^2\, \\
&&f=1-\frac{M}{\lambda}e^{-2\lambda r}\hh \psi=e^{\lambda r}\, .
\EE
This metric describes a $2D$ black hole in an asymptotically flat spacetime. Its horizon is located at
$r_\inds{H}=\frac{1}{2\lambda}\ln\frac{M}{\lambda}$,
and its spacetime curvature is
\ba
R=-\frac{\partial^2 f}{\partial r^2}
=4\lambda M \,e^{-2\lambda r} .
\ea
It takes value $R=4\lambda^2$ on the horizon. The Killing vector $\ts{\xi}=\partial_t$ is time-like in the black hole exterior, and it is space-like inside the black hole.

To cover a complete spacetime of the $2D$ black hole we introduce coordinates similar to the standard Kruskal coordinates. For this purpose we introduce null coordinates
\be
U=-\sqrt{\lambda\over M} \exp(-\lambda u),\
V=\sqrt{\lambda\over M} \exp(\lambda v)\, .
\ee
where
\be  \label{dsuv}
u={t-r_*}\hhh v={t+r_*}\hh
r_*=\int {dr\over f}={1\over 2\lambda}\ln \left(\exp(2\lambda r)-{M\over \lambda}\right) \, .
\ee
In these coordinates
\be \n{UV}
ds^2=-{1\over \lambda^2}{dU\, dV\over 1-UV}\hh
-\infty<U<\infty\, ,\quad -\infty<V<\infty\, ,\quad UV<1\, .
\ee

To construct the Carter-Penrose conformal diagram for the $2D$ metric  we introduce new null coordinates $(p,q)$
\be
U=\tan p \, ,\quad V=\tan q \, ,\quad \left(-{\pi\over 2}<p<{\pi\over 2} \, ,
 -{\pi\over 2}<q<{\pi\over 2}\right) .
\ee
In these coordinates the metric takes the form
\be\n{pqpq}
ds^2=-{ 1\over \lambda^2}{ dp\, dq\over \cos p\, \cos q\, \cos (p+q)}\,  .
\ee
Lines $p+q=\pm {\pi \over 2}$ correspond to spacelike curvature  singularities.

\begin{figure}[!hbt]
    \centering
    \vspace{10pt}
    \includegraphics[width=0.5 \textwidth]{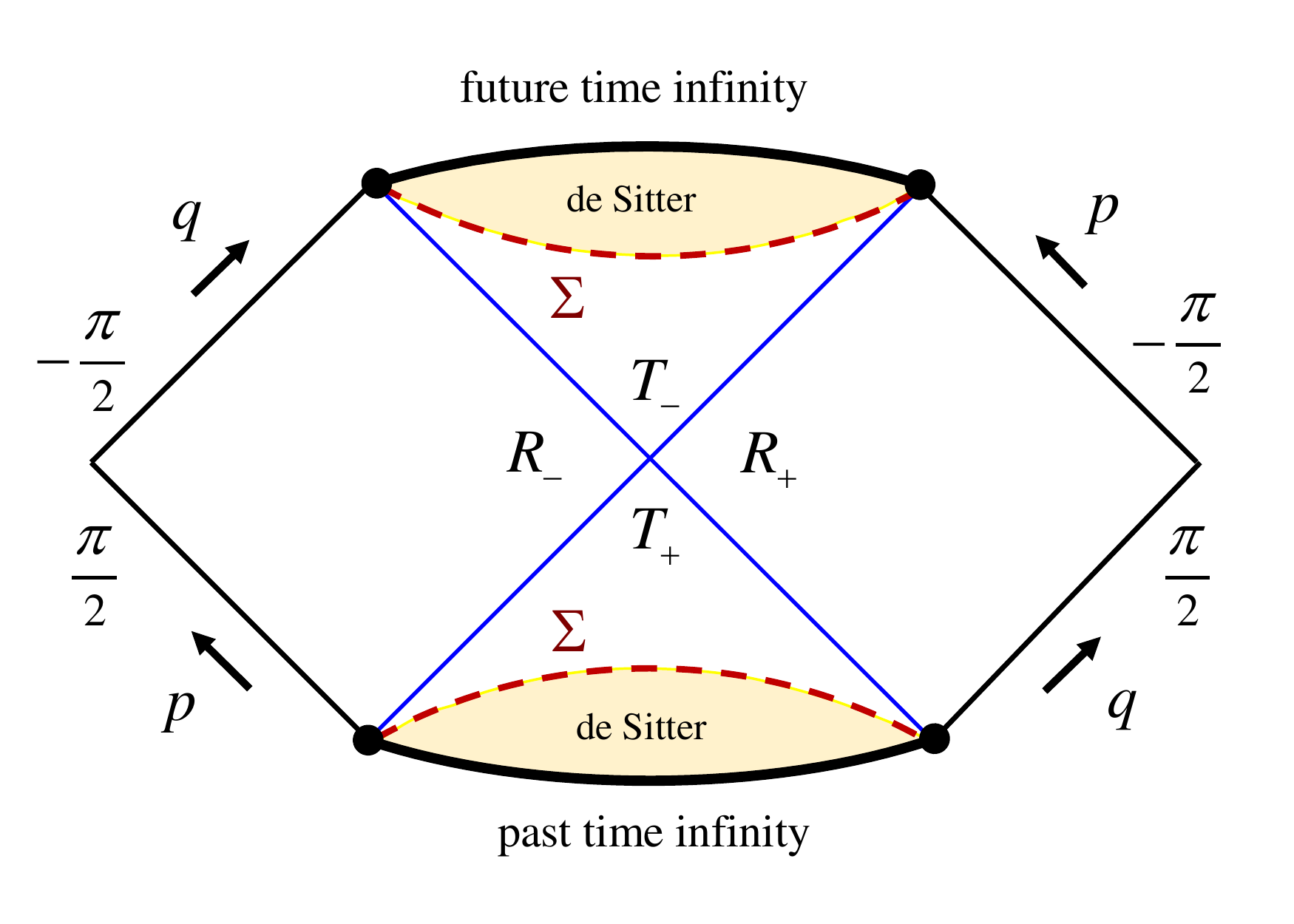}
    \caption{Conformal diagram for the  spacetime of the solution of LCG theory which consists of  an `eternal' $2D$ black hole domain glued to the de Sitter spacetimes. Dashed lines represent junction surfaces where the curvature reaches its maximal value $R=\Lambda$.}
    \label{2DBH}
\end{figure}

At line $r=r_{\Lambda}=-(1/2\lambda)\ln(\Lambda/(4\lambda M))$ the curvature $R$ reaches its critical value.
We impose a condition $\Lambda/(4\lambda^2)>1$, so that $r_{\Lambda}<r_H$ and the transition point is located inside the black hole.
Beyond this point the supercritical regime starts. In this regime the constraint equation takes the form $R=\Lambda$. The spacetime metric in this supercritical domain is locally de Sitter one. Analysis of the dilaton and gravitational equations shows that the control function $\chi$ does not vanish there. This means that after the solution enters its supercritical phase from the subcritical one it remains in this phase forever. A detailed discussion of gluing sub- and supercritical solutions and global structure of the $2D$  black holes in the LCG theory can be found in \cite{FZ1}. Figure~\ref{2DBH}  from this paper shows a conformal diagram for an eternal $2D$  black hole.

Similarly to the eternal $4D$ Schwarzschild black hole it has two domains $R_+$ and $R_-$ with asymptotically flat infinities. Regions $T_-$ and $T_+$ describe interiors of black and white holes, respectively. Instead of singularities these domains contain expanding and contracting de Sitter-like cores. As usual, such an eternal spacetime solution can be used to construct solutions for different physically interesting cases. For example, when a black hole is formed in gravitational collapse of a body, a boundary of this body is represented by a timelike curve on the diagram~\ref{2DBH}. A spacetime domain of the eternal black hole located to the left from this line should be cut and substituted by the metric inside the collapsing body. Let us summarize: In the LCG model of a $2D$ black hole its interior contains expanding de Sitter core instead of the singulatity.

\section{Four-dimensional spherically symmetric black holes in LCG theory}

Study of the interior of four-dimensional black holes in LCG theory is much more complicated \cite{FZ3}. One of the reasons is that the number of independent curvature invariants for spherically symmetric metrics is 4 instead of one scalar curvature invariant in $2D$ case.

When the curvature invariants are smaller than their critical value, a solution is subcritical and coincides with the Schwarzschild metric
\be \n{SCH}
ds^2=-\left(1-{2M\over r}\right) dt^2+{dr^2\over 1-{2M\over r}}+r^2 d\omega^2\, .
\ee
We assume that a transition to the supercritical regime takes place inside the black hole where the metric takes the form
\be \n{INTS}
ds^2=-{dr^2\over f}+f dt^2 +r^2 d\omega^2\hh f={2M\over r}-1\, .
\ee
In the black hole interior $f>0$ and introducing a new coordinate $\tau$
\be
\tau=\sqrt{r(2M-r)}+M \arcsin({M-r\over r})+{1\over 2}\pi M\, ,
\ee
one can write (\ref{INTS}) in the form
\ba\n{sm}
ds^2=-d\tau^2+B^2 dt^2+a^2 d\omega^2 \, ,
\ea
where $a$ and $B$ are functions of proper  time $\tau$
\be
B^2=f(r(\tau))\hh a=r(\tau)\, .
\ee
If the transition from sub- to supercritical regime takes place at $\tau=\tau_0$ than after this time the corresponding supercritical solution still has the same form. However, the metric functions $a(\tau)$ and $B(\tau)$ are different. They, as well as the control function $\chi$ should be determined by solving the corresponding set of equations. The form of these equations depends on how many inequality constraints are imposed and what is their structure.

The Riemann curvature tensor for the metric (\ref{sm}) has four independent non-vanishing components
\be \n{RIN1}
R_{\hat{\tau}\hat{t}\hat{\tau}\hat{t}}=-v\hh R_{\hat{\tau}\hat{\theta}\hat{\tau}\hat{\theta}}=-q\, ,\hh R_{\hat{t}\hat{\theta}\hat{t}\hat{\theta}}=u\hh R_{\hat{\theta}\hat{\phi}\hat{\theta}\hat{\phi}}=p\, .
\ee
The hat over the indices  means that the  components of this tensor are calculated in the orthonormal tetrad formed by unit vectors along the corresponding coordinate lines.
The independent basic curvature invariants $(p,q,u,v)$  are
\be \n{pquv}
p=\frac{\dot{a}^2+1}{a^2}\hh q=\frac{\ddot{a}}{a}\hh u=\frac{\dot{a}\dot{B}}{a B}\hh v=\frac{\ddot{B}}{B} .
\ee
A dot in these expressions means the derivative with respect to $\tau$.
For the subcritical Schwarzschild metric these curvature invariants are
\be\n{TRAN}
p=v=-2q=-2u={2M\over r^3}\, .
\ee

Invariant $p$ is positive definite, while the other invariants do not have a definite sign. This property makes the analysis of required inequality constraints more complicated. In paper \cite{FZ3} we discussed a case of constraints which are linear functions of curvature invariants (\ref{pquv}). The main results of this study are the following. The required property of the limiting curvature can be achieved if the following two constraints are imposed
\BE
&&\Phi_1=v-\Lambda=0 \, ,\n{VV}\\
&& \Phi_2=p-\mu q-\lambda \Lambda=0\, . \n{PQ}
\EE
Here $\Lambda=1/\ell^2$ is the limiting curvature parameter, while $\mu\in (0,1)$ and $\lambda$ are two dimensionless parameters.
The transition from sub- to supercritical regime happens at $\tau=\tau_0$ where $r=r_0=(2M\ell^2)^{1/3}$. For $\lambda>1+\mu/2$ the first constraint which is saturated is $v=\Lambda$. For a supercritical solution in this regime the invariants  $p$ and $|q|$ still grow. When condition (\ref{PQ}) is met the second constraint becomes active. This happens at some time $\tau_1\ge \tau_0$. Study of the behavior of the control function $\chi_1$ associated with constraint (\ref{VV}) shows that it does not vanish at this point. This means that at the second phase, that is beyond this point, both constraints (\ref{VV}) and (\ref{PQ}) are valid. In a general case such a supercritical solution does not leave this second phase after it enters it.

Constraint  (\ref{VV}) has a simple solution
\be \n{BBB}
B(\tau)=B_0 \cosh\left[\ell^{-1}(\tau-\tau_0+\phi)\right]\, .
\ee
Parameters $B_0$ and $\phi$ are defined by continuity conditions at $\tau=\tau_0$.
Solution (\ref{BBB})  has the same form for both supercritical phases. Let us consider a $(\tau,t)$ sector of the metric (\ref{sm}). Its $2D$ metric is
\be \n{2D}
dS^2=-d\tau^2+B^2(\tau) dt^2\, .
\ee
For $B(\tau)$ defined by (\ref{BBB}) this is $2D$ de Sitter metric with curvature ${}^{(2})R=\ell^{-2}=\Lambda$.
Thus, the supercritical solution for the interior of $4D$ black hole in its $2D$ sector $(\tau,t)$ is the same as a solution for the interior of $2D$ black hole discussed in the previous section.

The second constraint (\ref{PQ}) gives a nonlinear second order ordinary differential equation for function $a(\tau)$. For the imposed condition $0<\mu<1$ a solution of this equation has two turning points $a_{min}<a_{max}$, where $\dot{a}=0$. If $M\gg \ell$, then $a_{min}$ is close but slightly bigger than $\ell$, while $a_{max}\sim (2M\ell^2/\lambda)^{1/3}$.  Let us denote
\be
d\Gamma^2=-d\tau^2+a^2(\tau) d\omega^2\, .
\ee
Then the solution of  the constraint equation (\ref{PQ}) describes a $3D$ oscillating closed universe.

To summarize, the interior of a $4D$ spherically symmetric black hole in the LCG model with linear in curvature constraints is a $3D$ space which is exponentially expanding in one direction, determined by its Killing vector $\ts{\xi}=\partial_t$, and it is oscillating in the  other two (spherical) directions. It is possible to check that all the invariants $(p,q,u,v)$ for this solutions remain bounded and their values are uniformly restricted by a quantity proportional to the limiting curvature $\Lambda$. Derivation of these results and their detailed discussion can be found in \cite{FZ3}.

\section{Bouncing cosmologies in LCG theory}

As the last example of application of the LCG theory let us discuss a case of the cosmology. Namely, we assume that a homogeneous isotopic universe filled with thermal radiation is in its contraction phase. Its metric is
\be \label{COSM}
ds^2=-d\tau^2+a^2(\tau) d\gamma^2 \, ,
\ee
where $d\gamma^2=\gamma_{ij}dx^idx^j$ is a line element on a unit round 3D sphere $S^3$.
The metric $ds^2$  is conformally flat so that its Weyl tensor vanishes. We consider curvature invariants which are scalar functions of the Ricci tensor $R_{\alpha\beta}$.

According to the General Relativity  a collapse of such universe ultimately causes the cosmic scale factor $a(\tau)$ to reach zero and the contraction of the universe ends by the Big Crunch with the curvature singularity formation. In paper \cite{FZ2} we demonstrated that in the LCG model such a singularity is absent. Instead of it the universe has a Big Bounce after which it becomes expanding. Let us briefly discuss these results.

The Ricci tensor calculated for the metric (\ref{COSM}) is
\be
R^{\mu}_{\nu}=\mbox{diag}(3q,q+2p,q+2p,q+2p)\, ,
\ee
where
\be
q=\frac{\ddot{a}}{a}\hh p=\frac{\dot{a}^2+1}{a^2} .\n{ppqq}
\ee
Scalar curvature invariants are functions of $p$ and $q$, and the corresponding  inequality constraint can be written in the form
\be
\Phi=\Phi(p,q,\Lambda)\le 0\, .
\ee
The simplest choice of the constraint function $\Phi$ is
\be\n{LINCON}
\Phi=p-\mu q -\Lambda=0\hh 0<\mu<1
\, .
\ee
Let us note that this constraint is similar to the constraint (\ref{PQ}) which was discussed in the previous section. We again define a critical length by $\ell$, so that $\Lambda=\ell^{-2}$.

In the subcritical regime where $p-\mu q -\Lambda<0$ a solution coincides with a solution of the Einstein equations. The stress-energy tensor of the thermal radiation is
\ba\nonumber
T^{\mu}_{\nu}&=\mbox{diag}(-\varepsilon, P,P,P)\, ,
\ea
where  $\varepsilon$ is the energy density and $P$ is the pressure.
The conservation law $T^{\mu}_{\ \nu ;\mu}=0$  is satisfied if
\ba\label{Ca}
\varepsilon=A a^{-4} ,
\ea
where the factor $A$ is defined by the total (conserved) entropy of the universe $S$
\be \n{CCCC}
A=\nu \hbar c \Big( \frac{S}{k_\ins{B}}\Big)^{4/3}
\hh
\nu={3\over 16\pi^3}\Big( \frac{90}{n \pi}\Big)^{1/3}.
\ee
For pure electromagnetic radiation $n=2$ and  $\nu\approx 0.015 $.

At the contraction phase of the radiation dominated universe the following relation is valid
\be \n{pqE}
q=-p\, .
\ee
During the contraction of the universe both invariants $p$ and $|q|$ grow until the constraint function $\Phi$ reaches its critical value. After this a solution becomes supercritical and it follows the constraint equation (\ref{LINCON}). It is possible to show that for this solution the invariant $p$ monotonically grows until $a(\tau)$ reaches its minimal value $a_{min}$ which is of order of $\ell$ but slightly bigger than it. After this  the scale function $a(\tau)$ grows. Similarly to the case of the $4D$ black hole interior, there also exists a second turning point where $\dot{a}=0$ and $a(\tau)$ has maximum. However, before this the supercritical solution crosses the line (\ref{pqE}) on the $(p,q)$ plane. Because of the symmetry of the equation for the control function $\chi$ it vanishes at this point, and instead of further motion along the constraint (\ref{LINCON}) the solution returns to its subcritical regime. This subcritical solution describes an expanding universe filled with the thermal radiation. Thus in the LCG theory a contracting homogeneous isotropic universe  has a Big Bounce instead of a Big Bang singularity.   Derivation of these results and their detailed discussion can be found in a recent paper \cite{FZ2}.

\section{Discussion}

In this paper we discussed two types of models which satisfy the limiting curvature condition. In section~2 a wide class of spherically symmetric metrics  describing nonsingular black holes is presented. These metrics which contain two arbitrary functions of one variable  present a far-going generalization of Hayward-Vaidya metrics. One of the functions, a mass function $m(v)$, controls the black hole mass and the null fluid flux responsible for its change.
The other one, a structure function $F(x)$, is responsible for properties of the nonsingular black hole interior.
These metrics have an interesting property: The  curvature invariants calculated for them do not depend on the mass function and after imposing simple restrictions of the structure function they obey the limiting curvature condition.

In the second part of the paper we discuss a new recently proposed limiting curvature theory of gravity. In this theory the Einstein-Hilbert gravitational action is modified by adding to it terms which guarantee that inequality constraints for chosen curvature invariants are satisfied. We discussed application of LCG theory to three cases: $2D$ dilaton gravity black holes \cite{FZ1}, $4D$ spherically symmetric LCG modification of the Schwarzschild metric \cite{FZ3} and to bouncing cosmologies \cite{FZ2}. As further applications of the LCG theory it would be interesting to discuss such ``hot" subjects as the large scale structure formation in bouncing cosmological LCG models, possible anisotropy suppression in anisotropic cosmologies and the  interior structure of charged and rotating $D$ black holes.

\section*{Acknowledgments}

The author is grateful to  the Natural Sciences and Engineering Research Council of Canada
and the Killam Trust for their financial support. He also thanks Andrei Zelnikov for useful remarks.

\vfill



\begin{thebibliography}{0}
\bibitem{MA1}\BY{Markov, M.A.} \IN{JETP Letters}{36}{1982}{266}.
\bibitem{MA2}\BY{Markov, M.A.} \IN{Annals Phys.}{155}{1984}{333}.
\bibitem{Hayward}\BY{Hayward, S.A.}\IN{Phys. Rev. Lett.}{96}{2006}{031103}.
\bibitem{F1} \BY{Frolov, V. P.} \IN{JHEP}{1405}{2014}{049}.
\bibitem{B1}\BY{Bardeen, J.M. } preprint gr-qc 1406.4098, {2014}.
\bibitem{F2} \BY{Frolov, V. P.} \IN{Phys. Rev. D}{94}{2016}{104056}.
\bibitem{B2}\BY{Bardeen, J.M.} preprint gr-qc 1811.06683, {2018}.
\bibitem{FZ1} \BY{Frolov, V. P. \atque Zelnikov, A.} \IN{JHEP}{08}{2021}{154}.
\bibitem{FZ3} \BY{Frolov, V. P. \atque Zelnikov, A.}
\TITLE{Spherically symmetric black holes in the limiting curvature theory of gravity},
preprint gr-qc 2111.12846, {2021}.
\bibitem{FZ2} \BY{Frolov, V. P. \atque Zelnikov, A.} \IN{Phys. Rev. D}{104}{2021}{104060}.

\end{thebibliography}

\end{document}